# An examination of applicability of face recognition sensors in public facilities

Takuji Takemoto[1], Takashi Ota[2,] and Hiroko Oe*[3]

[1] *University of Fukui, 3-9-1 Bunkyo, Fukui 910-8507, Japan*

[2] *OMRON Corporation, 2-2-1 Nishikusatsu, Kusatsu 525-0035, Japan*

[3] *Bournemouth University, 89 Holdenhurst road, Bournemouth, BH8 8EB, UK*

\* Corresponding author. E-mail address: hoe@bournemouth.ac.uk.

**Abstract**

**Purpose:** This study aimed to explore the usability and applicability of face recognition sensors in public spaces to collect customer footfall data, which could then be analysed and evaluated for facility design and planning

**Methodology:** Nine OMRON sensors were provided for the project and installed at five locations in a public facility for three months. The project was carried out by a local consortium with the cooperation of local technology-based Small Medium sized Enterprises (SMEs), business organisations, and a local university. Collected data was analysed using data-mining software to develop a result report with diagrams, and reveal issues and potential for practical application in the future.

**Findings:** It has been found that this technology could be applied for further consumer behavioural analysis, for example, analysing the relationship between product displays and purchasing behaviour, or looking at the link between consumers' attributes and their buying behaviour. Moreover, the collected data can be further studied to develop a more detailed analysis of the relationships between the data collected from different points of installation.

A critical issue found was about how to protect the privacy of the people whose data the sensors collected (i.e., image rights, and other privacy-related issues), which suggests the need for guidelines on ethical data collection and raises questions on how to get agreement from potential participants in the experiment.

**Implication and limitation:** Although it was acknowledged that this project remained at pilot level and would need to expand before more robust implications and recommendations could be developed, the experimental outcome suggests that face recognition sensors have the potential for commercial use. Collecting and analysing customers' behavioural data can contribute to marketing strategy and planning. The study also discusses the necessity of enhancing business opportunities through open innovation, in this case based on a consortium inviting local technology-oriented SMEs, universities, and other stakeholders to support the local economy. The implications of this study could inspire others to start new businesses and to support the local economy and small enterprises.

# 1. Introduction

*1.1 Background of the study*

Many companies are interested in learning what business opportunities can be enhanced with the implementation of the possibilities offered by the of Internet of things (IoT) and artificial Intelligence (AI) (Toniolo et al. 2020). The IoT involves not just products and the Internet, but also the handling of enormous amounts of data, and the key is how to utilise that big data (An et al., 2019; Dai et al., 2019).

As the Fourth Industrial Revolution (4IR) has progressed, new digital technologies have given rise to unprecedented new business models. The manufacturing industry, especially, has used IoT to develop new services and industrial sectors. It is necessary to respond to drastic changes in the industrial structure with technological innovations. This movement provides new customer value that exceeds the scope of conventional manufacturing, with innovations such as Mobility as a Service (MaaS). Intersectional collaborations between a range of different sectors have accelerated this drastic change to modern industry (e.g., Kobayashi et al. 2019; Soh & Connolly, 2020). In fact, MaaS, a new concept of mobility, has the potential to transform existing cities into smart cities with the implementation of IoT and AI as the basis for operations (Nikitas et al., 2020; Rayes et al., 2020).

*1.2 Outline of the experiment*

In this study, we used face recognition sensors, developed by OMRON, to evaluate the operational issues and applicability of practical data collection. These sensors have two functions. The first is a customer base analysis function that recognises human faces and attributes, which are captured on camera and recorded with time stamps (e.g., estimated age and gender). The second is a heat map function that divides the area covered by the sensor into a grid composed of six by eleven smaller squares, and counts the number of positions where faces are recognised. Eiheiji Town, Fukui Prefecture agreed to participate in this project.

**2. Method**

*2.1 Public announcement*

To publicise this project and make potential participants aware of the data collection via face sensors, we hosted an exhibition at Fukui Prefecture Trade Fair prior to the launch of the experiment. During this public announcement, we also gathered visitors' opinions regarding the potential of a collaborative research project based on a consortium of relevant stakeholders (**Figure 1**).

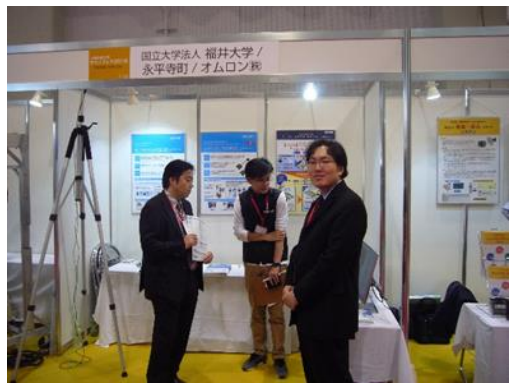

**Figure 1. The booth for this experiment at Fukui Prefecture's trade fair**

For the three-month period of the experiment, sensors were placed at three entrances of the town hall (sensors A–C), one at the entrance to a public bathhouse (sensor D), and five in the bathhouse's souvenir shop (sensors E–I). In total, nine sensors were installed. Due to concerns regarding the image rights of individuals, we decided to indicate clearly beforehand, by means such as posters, that sensors were to be installed, and clearly displayed the capture area for data collection so that those who did not want to their data to be collected could avoid being involved in this experiment (**Figure 2**).

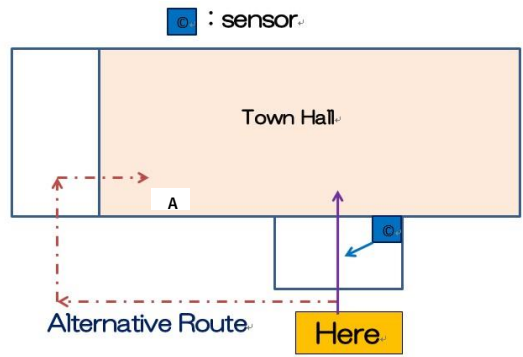

**Figure 2. Notice of 'how to avoid being recorded' with an alternative route**

*2.2 Outline of the experiment*

We arranged the thematic targets based on sensor location type. Sensors A–F were used for customer behavioural analysis (six sensors), and sensors G–I were used for heat map analysis (three sensors). To ensure operability, we also set up internet access via portable wi-fi devices each location.

The places where the sensors were installed, and the heights at which they were installed, were as follows:

A, B, and C, Eiheiji Town Hall entrances: ~1.4 m.

D, entrance to the public bathhouse: ~1 m.

E, on top of a ticket vending machine in the souvenir shop restaurant: ~1.7 m.

F, above a cash register in the souvenir shop: ~2 m.

G, above the refrigerated display shelves in the souvenir shop: ~2 m.

H and I, on pillars in the souvenir shop: ~2 m.

As an example, **Figure 3** shows the sensor located at point E.

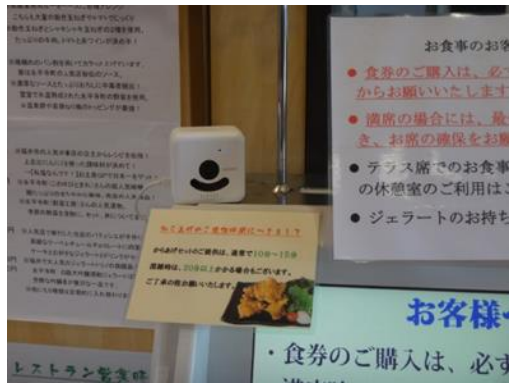

**Figure 3. Sensor E, on top of a ticket vending machine in the restaurant of the souvenir shop (~1.7 m)**

The data collected for the two thematic categories, 'customer behavioural analysis' and 'heat maps' over the three-month period were summarised and analysed using Excel. The accuracy of the data was corroborated in interviews with governmental officials from the local council, as well as discussions with the staff of the souvenir shop regarding sales records and customers' purchasing behaviour.

## 3. Findings and analysis

*3.1 Overall findings*

Due to obstacles placed in front of the sensors, power outages (and internet connection difficulties resulting from the outages), and the failure of voltage adapters, there were some days on which no customer base data could be collected. Measurements for heat-mapping analysis, however, were carried out without any failures and the overall data obtained was good enough for analysis. The results and analyses are summarised in the next sections.

*3.2 Location A*

This sensor mainly recorded visitors to the town hall. Although there was a slight increase in the number of visitors at the beginning of the year, the average number of data captures was stable. When the data were viewed by day of the week, the number of visitors on Thursdays was notably low; however, there were no significant differences between the numbers of visitors on other weekdays. **Figure 4** shows a data outline of visitor flow in one week; visitors are grouped by demographic.

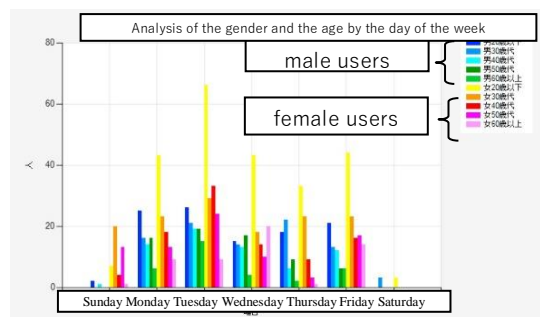

**Figure 4. Data collected from Sensor A.**

*3.3 Location B*

Sensor B was located in the town hall building's library annex. Fewer visitors were detected on Mondays, when the library is closed. The number of visitors was evenly distributed over weekdays when the library is open. An increase in visitors was shown during the weekend, when the library is also open (**Figure 5**).

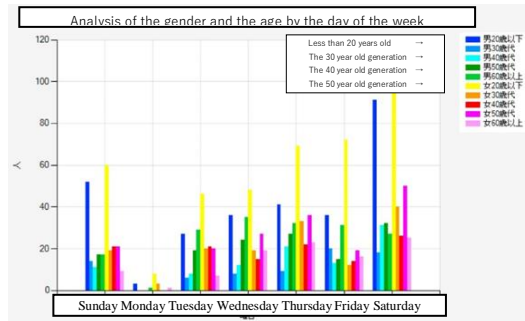

**Figure 5. Data collected from Sensor B**

*3.4 Location C*

When compared to the number of visitors recorded by employees of the town hall, the number of faces recognised by Sensor C was smaller than anticipated, even after subtracting the assumed number of people taking alternative routes to avoid being captured (**Figure 6**). This outcome implies that it is necessary for the researchers to evaluate the data collection environment to validate the figures. The implementation of measures to ensure a reliable dataset is essential.

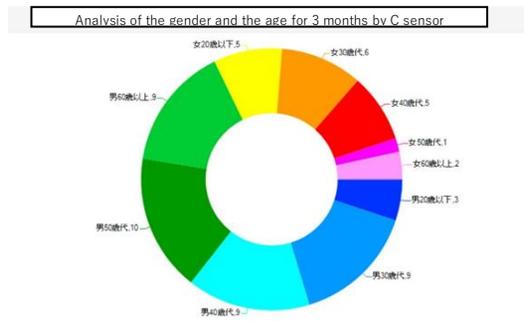

**Figure 6. Data collected from Sensor C (recognised faces)**

*3.5 Location D*

Mainly because some obstacles that disturbed data transmission were placed in front of the sensor, Sensor D was unable to collect data for a few days of the experiment's duration. In addition, as a result of backlighting, some faces could not be recognised (**Figure 7**). Location D is one of the busiest areas, near the public bathhouse and the souvenir shop. The amount of data collected was much smaller than we expected. As with the outcome of Location C, this result again indicates the importance of ensuring reliable data collection.

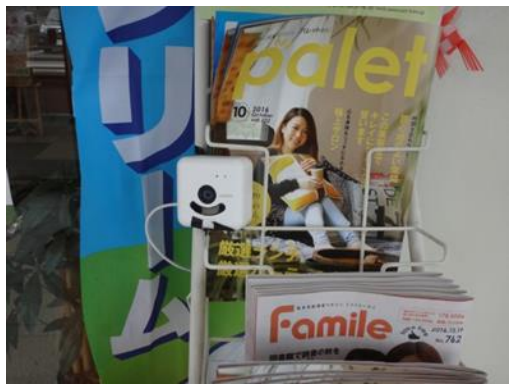

**Figure 7. An image of the location of Sensor D (placed on a magazine rack)**

*3.6 Location E*

Because sensors E–I were placed not at entrances but indoors, they collected data in a reliable manner. The totals for E showed a clear decline from November to December and from December to January. When staff were interviewed for the purposes of evaluating the collected data, they explained that when the weather turned cold and snowy, the number of visitors declined significantly (**Figure 8**). To analyse the data and develop more robust conclusions, more detailed data profiling and examination of seasonal activities is required.

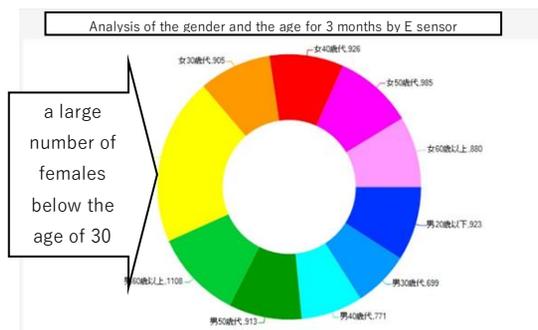

**Figure 8. Data collected from Sensor E: A large number of females below the age of 30 are captured**

*3.7 Location F*

There were clear fluctuations from day to day in the number of people detected by this sensor (**Figure 9**). The sensor was located near the souvenir shop, and the shop staff explained that the amount of data collected depended on the cash register they used. The sensor was positioned so as to capture the customers who used two particular cash registers, and the sensor mostly recognised their faces.

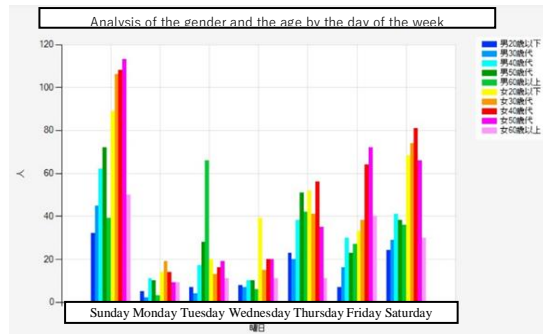

**Figure 9. Data collected from Sensor F**

*3.8 Location G*

**Figure 10** shows the six-by-eleven grid format of the data capture. Each square was numbered, e.g., 1A, 2B, 3C. We tracked the data by time of day and found that the number of visitors in square 3B was highest just before the shop closed; we believe this was because the sensor was close to the door and captured people as they left. In addition, the data showed the highest number of people on Saturdays and Sundays, and slightly above-average numbers on Thursdays and Fridays. There were few measurements for the product shelves near the sensor, most likely because the sensors only recognise faces from the front. Another technical issue found was that the sensor recognised a substantial number of visitors' faces in square 3E late at night on two particular days; this was most likely a data collection error and must be investigated further ensure the validity of our results.

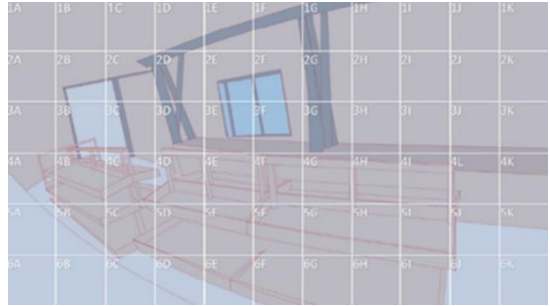

**Figure 10. Data collect format of Sensor G: 6×11 matrix**

*3.9 Sensor H*

This sensor captured most data in square 3C, which might have detected the visitors who were waiting in a queue (**Figure 11**). However, it is believed that the number of people looking away from the sensor resulted in a decrease in the number of recognised faces. The restaurant's menu is displayed on the wall, which could be the reason why visitors' faces are turned away from the sensor.

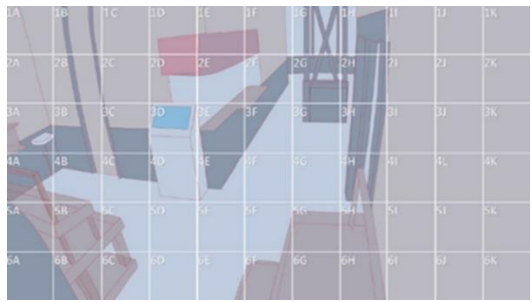

**Figure 11. Data collection format of Sensor H: 6 × 11 matrix**

*3.10 Sensor I*

A large amount of data was collected from Sensor I, which was positioned near cash registers and notice boards. Square 2F captured the largest portion of data; we hypothesised that the reason for this is that popular souvenirs with local area-specific designs were located in this position. The data collected from square 5E also indicated unique customer footfall, with a greater number of faces recognised at a certain time of day. **Figure 12** shows the data collection format.

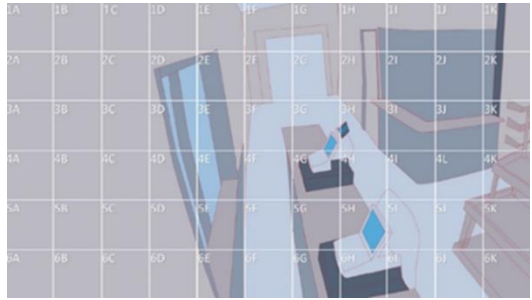

**Figure 12. Data collection format of Sensor I: 6 × 11 matrix**

**4. Discussion**

*4.1 Ethical requirements*

This experiment was carried out during the three-month period as planned and the data collection went well overall but, as noted, some issues did arise. During the preparatory stage within the consortium, the critical theme was how to meet privacy requirements as the collected data includes facial recognition data. Even with the public announcement of the event and the explanation to local residents through the media and notice boards, it was not realistic for the consortium to collect individual consent forms from every participant in the experiment. Researchers are now looking into how to blur images of faces to secure the participants' privacy while still collecting minimal demographic data (e.g., age band and gender).

As a basic ethical condition, a detailed announcement will be included in journal articles, books, and any other publication featuring this study, including online sources. The announcement will state that all captured data is fully anonymised and to be used only for experimental purposes in the domains of marketing and technology. Furthermore, it has been made clear that the visitors' movements and faces were not recorded and all data was in a quantitative format, to be used for statistical analyses only.

*4.2 Technological enhancement and the impact of open innovation*

The other significant issues that presented themselves during this study centred around technical difficulties. To avoid the incorrect collection of data under certain conditions (e.g., backlighting, mirrored figures in the windows), technical solutions are required.

As Kumar et al. (2019) discussed, with advancements in security, intelligent facial recognition systems could be adapted to navigate services based on the IoT. Similarly, Sagar and Narasimha (2019) suggested that analysis of facial recognition data has the potential to contribute to the social economy. The outcomes of this research suggest that open collaborative data sharing among the relevant stakeholders should be the basis for future open innovation in the community. Indeed, this could represent the next major frontier in the field of software engineering, both in research and in practice (Runeson, 2019).

*4.2 Further research opportunities*

This experimental study is a community-based experiment conducted by a consortium of local council members, the local university, and the scientific organisation OMRON. Privacy requirements were found to be a critical theme in continuing this detailed experiment to collect data to improve and expand the applicability of the sensors to real-life practices. An internet connection is essential for the operation of sensors; without it, stable collection of data is not possible. Obstacles between visitors and the sensors also interrupt the smooth transmission of data.

The analysis of heat maps obtained throughout this study displayed only the temperature in each area. To develop more actionable conclusions from this particular function of the sensors, the heat maps could be linked to thermostats in order to automatically adjust to the ideal room temperature.

If the automatically collected data can be applied to customer behavioural analysis, the use of sensors could contribute to designing marketing strategies for the local shops and facilities where the sensors are installed (Samadiani et al., 2019). Sensor technologies could be also applied to the health sector to monitor the flow of patients' and ensure their safety (García-Peñalvo & Franco-Martín, 2019). For this study, the raw data was analysed using a basic Excel spreadsheet. However, more advanced data handling software would be more suitable for dealing with such a great quantity of data (Priyadarshini et al., 2020).

This project should be extended so that local businesses can make use of the sensors and other assistive technologies to maintain and improve their operations, which may facilitate the regeneration of the local economy (Lindgren et al., 2020). As a collaborative research consortium, the members of the team should use the project resources to develop harmonised business opportunities and bring about mutually beneficial relationships among community-based stakeholders.